\documentclass[aps,twocolumn,prl]{revtex4-1}
\usepackage{graphicx,amsmath, amsfonts,bm, color,amssymb}
\usepackage{gensymb,sidecap,indentfirst}

\usepackage{xcolor,hyperref}
\hypersetup{
   colorlinks,
   linkcolor={blue!50!black},
   citecolor={blue!50!black},
   urlcolor={blue!80!black}
}

\usepackage{enumitem,enumerate}
\renewcommand{\th}{^{\mbox{\tiny th}}}
\newcommand{\pa}{\partial}

\renewcommand{\=}{\!=\!}
\newcommand{\1}{^{\mbox{\tiny (1)}}}

\newcommand{\Gb}{\ensuremath{\overline{G}}}
\newcommand{\calBold}[1]{\mbox{\boldmath${\cal #1}$}}

\newcommand{\B}[1]{{\bm{#1}}}
\newcommand{\C}[1]{{\mathcal{#1}}}
\newcommand{\dbar}{{\,\mathchar'26\mkern-12mu d}}

\begin{document}

\title{Anisotropic Structural Predictor in Glassy Materials}
\author{Zohar Schwartzman-Nowik$^1$, Edan Lerner$^2$, and Eran Bouchbinder$^1$}

\affiliation{$^{1}$Chemical and Biological Physics Department, Weizmann Institute of Science, Rehovot 7610001, Israel\\
$^{2}$Institute for Theoretical Physics, University of Amsterdam, Science Park 904, 1098 XH Amsterdam, The Netherlands}

\begin{abstract}
There is a growing evidence that relaxation in glassy materials, both spontaneous and externally driven, is mediated by localized soft spots.
Recent progress made it possible to identify the soft spots inside glassy structures and to quantify their degree of softness. These
softness measures, however, are typically scalars, not taking into account the tensorial/anisotropic nature of soft spots, which implies orientation-dependent
coupling to external deformation. Here we derive from first principles the linear response coupling between the
local heat capacity of glasses, previously shown to provide a measure of glassy softness, and external deformation in different directions.
We first show that this linear response quantity follows an anomalous, fat-tailed
distribution related to the universal $\omega^4$ density of states of quasilocalized, nonphononic excitations in glasses. We then construct
a structural predictor as the product of the local heat capacity and its linear response to external deformation, and show that it offers enhanced predictability
of plastic rearrangements under deformation in different directions, compared to the purely scalar predictor.
\end{abstract}

\maketitle

{\em Introduction.---} At the heart of resolving the glass mystery resides the need to quantify the disordered structures inherently associated with glasses and to relate them to glass properties and dynamics, most notably spontaneous and driven structural relaxation~\cite{paddy_huge_review_2015,falk_review_2016}. Numerous attempts to address and meet this grand challenge have been made~\cite{Schober_correlate_modes_dynamics,widmer2008irreversible,manning2011,thermal_energies,Barrat_pre_2009_local_elasticity,tanguy2010,tanaka_PRX_2018,manning_defects,machine_learning_1,machine_learning_2,machine_learning_3,machine_learning_4,falk_prl_2016,Falk_prl_2018_art,Falk_local_yield_stresses_PRE_2018, plastic_modes_prerc,Ding2014}, aiming at defining structural indicators with predictive powers. Achieving this goal would constitute major progress in understanding glassiness and would provide invaluable insight for developing macroscopic theories of deformation and flow of glasses.

Recently accumulated evidence suggests that spatially localized soft spots are the loci of glassy relaxation, and hence are highly relevant for glass dynamics. These localized soft spots have been related to quasilocalized, nonphononic excitations in glasses~\cite{widmer2008irreversible,manning2011,thermal_energies}, whose universal $\omega^4$ density of states ($\omega$ is the vibrational frequency) has been also established recently~\cite{modes_prl_2016,protocol_prerc,modes_prl_2018,ikeda_pnas}. Among the structural predictors proposed, most relevant here is the normalized local thermal energy~\cite{thermal_energies}, which quantifies the inter-particle interaction contribution to the zero-temperature heat capacity, termed hereafter the local heat capacity (LHC) $c_\alpha$ ($\alpha$ is the interaction index).

The LHC $c_\alpha$ is a general (system/model-independent), first principles statistical mechanical quantity that reveals soft spots in glassy materials~\cite{thermal_energies}. Yet, the LHC is a scalar that quantifies the resistance to motion {\em in some unknown direction}. That is, like previously proposed structural predictors in glasses (with the exception of~\cite{falk_prl_2016,Falk_local_yield_stresses_PRE_2018,Falk_prl_2018_art}), the LHC misses important tensorial/anisotropic information about the coupling to deformation in a certain direction. For example, an extremely soft spot can be completely decoupled from external forces applied in a certain direction and hence irrelevant for the glass response in this direction.

In this Letter, we develop and quantitatively test a theory that allows to identify particularly soft glassy structures, explicitly revealing their anisotropic nature and their intrinsic coupling to the direction of externally applied forces. The theory is developed in two steps; first, the linear response coupling of $c_\alpha$ to external deformation tensors $\bm{\mathcal{H}}(\gamma)$, parameterized by a strain amplitude $\gamma$, is derived. The resulting quantity, $dc_\alpha/d\gamma$, is shown to follow an anomalous, fat-tailed distribution related to the universal $\omega^4$ density of states of quasilocalized, nonphononic excitations in glasses. Second, a structural predictor defined as the product $c_\alpha\,dc_\alpha/d\gamma$ is proposed and shown to filter out soft spots that are not coupled to external deformation of interest. Finally, a metric for quantifying the predictive power of structural predictors is proposed and extensive computer simulations are used to show that $c_\alpha\,dc_\alpha/d\gamma$ offers enhanced predictability
of plastic rearrangements under deformation in different directions, compared to the LHC $c_\alpha$ alone.

{\em Linear response coupling of the LHC to external deformation.---} The starting point for our development is the zero-temperature local heat capacity~\cite{thermal_energies,comment}
\begin{equation}
\label{eq:LHC}
c_{\alpha}\!\equiv\!\frac{1}{\tfrac{1}{2} k_B}\frac{\partial\!\left\langle\varphi_{\alpha}\right\rangle_{_T}}{\partial T}\bigg|_{T=0} \ ,
\end{equation}
where $\langle{\varphi_\alpha}\rangle_{_T}\!=\!\int\!{\varphi_\alpha}({\B x})\exp\!\left(\!{-\frac{{\C U}({\B x})}{k_B T}}\!\right)\!d{\B x}/\!\int\!\exp\!\left(\!{-\frac{{\C U}({\B x})}{k_B T}}\!\right)\!d{\B x}$, ${\B x}$ is a vector of the positions of all particles, $\varphi_\alpha$ is the potential energy of any pair of interacting particles, $\C U({\B x})\!=\!\sum_\alpha\!\varphi_\alpha$, and $k_B$ is Boltzmann's constant. The sum over the LHC, $\tfrac{1}{2} k_B\sum_\alpha\!c_\alpha\=\partial\!\left\langle\C U\right\rangle_{_T}\!/\partial T\big|_{T=0}$, is the thermodynamic, zero-temperature heat capacity $C_{\rm V}$.

An analytic low-temperature expansion of $\langle{\varphi_\alpha}\rangle_{_T}$ allows to explicitly calculate $c_\alpha$~\cite{thermal_energies}, which takes the form $c_{\alpha}\=\calBold{\varphi}_\alpha''\!:\!\calBold{M}^{-1}-{\bm f}_\alpha\!\cdot\!\calBold{M}^{-1}\cdot\calBold{U}{'''}\!:\!\calBold{M}^{-1}$, where $\cdot$ denotes a contraction over a single index of the relevant tensors and $:$ over two indices. A prime, here and hereafter, denotes a partial derivative with respect to $\bm x$, ${\bm f}_\alpha\=\calBold{\varphi}_\alpha'$ are frustration-induced internal forces, and $\calBold{M}\!\equiv\!\pa^2 {\C U}/\pa{\B x}\pa{\B x}$ is the Hessian matrix whose eigenvalues are $\omega^2$, where $\omega$ is a vibrational (normal mode) frequency. The low-frequency vibrational spectra of glasses feature, in addition to extended phononic excitations (long-wavelength plane-waves), also quasilocalized nonphononic excitations, sometimes termed soft glassy modes~\cite{modes_prl_2016,protocol_prerc,modes_prl_2018}. The latter follow a universal density of states (DOS) $D_{\rm G}(\omega)\!\sim\!\omega^{4}$ that is different from Debye's theory~\cite{kittel1996introduction}.

\begin{figure}[ht]
 \centering
 \includegraphics[width=0.85\columnwidth]{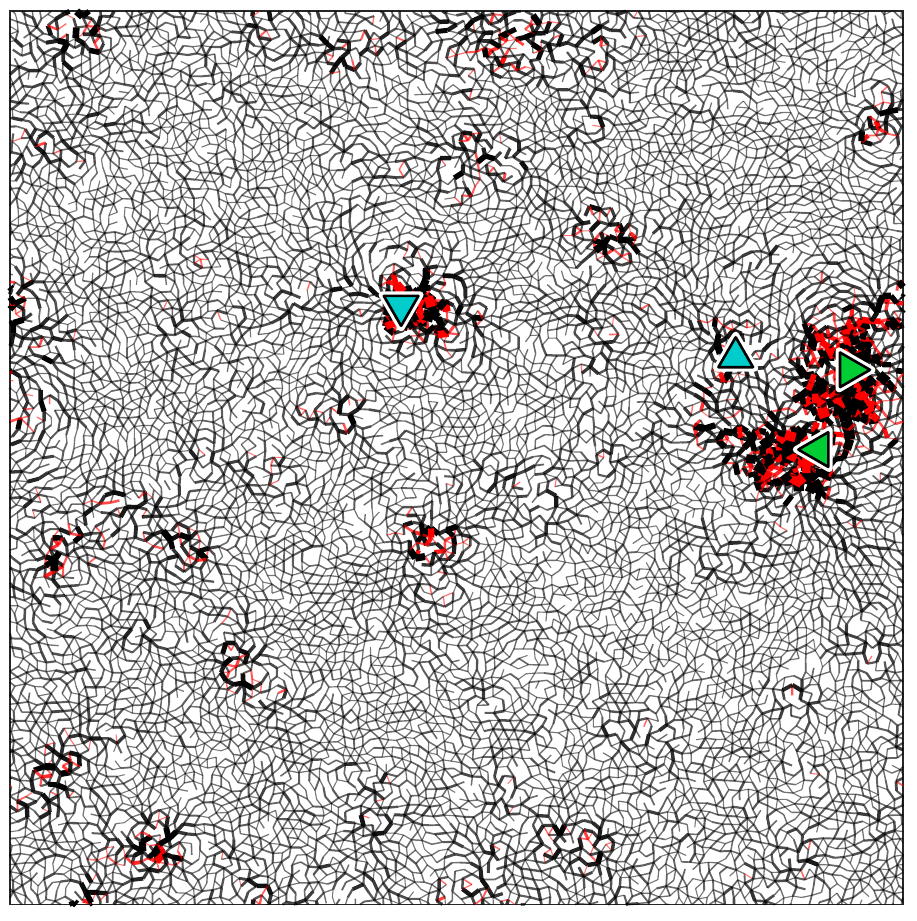}
 \caption{The local heat capacity (LHC) $c_\alpha$, cf.~Eq.~\eqref{eq:LHC}, for a glass composed of $N\!=\!10000$ particles~\cite{SM}. The magnitude of $c_\alpha$ is represented by the thickness of the lines connecting particles and black/red correspond to positive/negative values, see~\cite{SM} for details about the thresholding procedure employed. Regions with anomalously large $|c_\alpha|$, i.e.~soft spots, are clearly observed. The right/left triangles correspond to the first plastic events under positive/negative simple shear AQS deformation and the up/down ones to the first plastic events under positive/negative pure shear AQS deformation.}
 \label{fig:LTE}
\end{figure}
The LHC $c_\alpha$ is far more sensitive to soft quasilocalized modes than to extended phonons, i.e.~it filters out the effect of phonons, and is dominated by its second contribution, which is proportional to the frustration-induced internal forces ${\bm f}_\alpha$ and to $(\calBold{M}^{-1})^2$ (scaling-wise)~\cite{thermal_energies}. The spatial distribution of $c_\alpha$ reveals soft spots, see Fig.~\ref{fig:LTE}, which are highly correlated with the loci of plastic rearrangements under external driving (marked by the superimposed triangles, to be further discussed below).

The soft spots are characterized by a degree of softness determined by the typical magnitude of $c_\alpha$ in its vicinity (note that the local stiffness $\omega^2$ of the potential energy landscape scales as $c_\alpha^{-2}$), which quantifies the collective potential energy barrier that should be overcome in order to induce a structural rearrangement (the barrier is proportional to $\omega^6$ in the cubic approximation~\cite{Lacks_2006_barrier_scaling,plastic_modes_prerc}). The LHC $c_\alpha$, however, contains no information whatsoever about the {\em direction}, neither in the potential energy landscape of the glass nor in real space, in which the barrier is lowest. Consequently, an external driving force applied in a certain direction or a spontaneous thermal fluctuation that generates a local force in a certain direction, may or may not push a soft spot towards its activation barrier. In short, soft spots are expected to be tensorial/anisotropic objects that cannot be comprehensively described by scalar measures.

To demonstrate the tensorial/ansiotropic nature of soft spots, let us focus again on Fig.~\ref{fig:LTE}, where $4$ plastic rearrangement events are presented (triangles). Each of these correspond to the first plastic event of the same glass under external deformation applied in $4$ {\em different directions}. In particular, we applied Athermal Quasi-Static (AQS)~\cite{lemaitre2004_avalanches,lemaitre2006_avalanches,SM} simple and pure shear
$$\bm{\mathcal{H}}_{\rm simple}\=\left(\!\begin{array}{cc} 1 & \pm\gamma \\ 0 & 1 \end{array}\!\!\right)\quad\hbox{and}\quad\bm{\mathcal{H}}_{\rm pure}\=\left(\!\begin{array}{cc} 1\pm\gamma/2 & 0 \\ 0 & 1\mp\gamma/2 \end{array}\!\!\right)\!,$$ respectively, in both the positive/negative ($+/-$) directions, where $\gamma$ quantifies the amplitude of deformation. All $4$ plastic events occurred at soft spots, i.e.~regions of abnormally large LHC, but at $4$ {\em different} ones. This clearly demonstrates that soft spots are tensorial/anisotropic objects that feature different coupling to deformation in different directions.

To develop a theory that goes beyond the scalar LHC as a structural predictor in glasses, we set out to calculate the linear response coupling of the LHC to external deformation. That is, given a certain globally imposed deformation $\bm{\mathcal{H}}(\gamma)$, we aim at calculating analytically $dc_\alpha/d\gamma$ associated with $\bm{\mathcal{H}}(\gamma)$. The structure of the differential operator $d/d\gamma$ reflects the intrinsically disordered nature of glasses; it is composed of two contributions~\cite{lutsko}, one represents the affine response of ordered systems $\pa/\pa\gamma$ and the other represents the additional non-affine motions associated with disorder-induced forces, $-\calBold{U}^{\gamma}{'}\!\cdot\!\calBold{M}^{-1}\!\cdot\!\pa/\pa{\bm x}$, where the superscript $^\gamma$ is a shorthand notation for $\pa/\pa\gamma$ and $\calBold{U}^{\gamma}{'}$ are the mismatch forces (that drive the non-affine motions). Operating with $d/d\gamma\=\pa/\pa\gamma-\calBold{U}^{\gamma}{'}\!\cdot\!\calBold{M}^{-1}\!\cdot\!\pa/\pa{\bm x}$ on $c_\alpha$, we obtain to leading order in $\calBold{M}^{-1}$ (the complete and exact result, including all orders in $\calBold{M}^{-1}$, is presented in~\cite{SM})
\begin{eqnarray}
\label{eq:der_LHC}
& & dc_{\alpha}/\!d\gamma\!\simeq\!
-\,\calBold{U}^{\gamma}{'}\!\!\cdot\!\calBold{M}^{-1}\!\!\cdot\!\left(\calBold{U}{'''}\!\!\cdot\!\calBold{M}^{-1}\!\cdot\calBold{U}{'''}\!\!:\!\calBold{M}^{-1}\right)\!\cdot\!\left({\bm f}_\alpha\!\cdot\!\calBold{M}^{-1}\right)\nonumber\\
& &
-\!\left(\calBold{U}^{\gamma}{'}\!\!\cdot\!\calBold{M}^{-1}\!\!\cdot\calBold{U}{'''}\!\cdot\!\calBold{M}^{-1}\right)\!\!:\!\!\left(\calBold{M}^{-1}\!\cdot\calBold{U}{'''}\!\cdot\!\calBold{M}^{-1}\!\cdot\!{\bm f}_\alpha\right).
\end{eqnarray}

Equation~\eqref{eq:der_LHC} shows that the largest values of $dc_\alpha/d\gamma$ emerge from a fourth power of $\calBold{M}^{-1}\!\!\sim\!\omega^{-2}$ (scaling-wise), coupled to the energy anharmonicity tensor $\calBold{U}{'''}$, to the internal force vector ${\bm f}_\alpha$ and to the mismatch force vector $\calBold{U}^{\gamma}{'}$. Note that similarly to $c_\alpha$ (see expression above), the existence of frustration-induced internal forces ${\bm f}_\alpha$ --- an intrinsic signature of glassy disorder --- is essential for the emergence of abnormally large values of $dc_\alpha/d\gamma$. While the expression for $dc_\alpha/d\gamma$ (in Eq.~\eqref{eq:der_LHC} or its exact counterpart in~\cite{SM}) is universal, the specific information regarding the applied deformation $\bm{\mathcal{H}}(\gamma)$ for which the linear response is calculated is encapsulated in the partial derivative $\pa/\pa\gamma$~\cite{SM}, here through the mismatch force $\calBold{U}^{\gamma}{'}$. The validity of the analytic expression for $dc_\alpha/d\gamma$ has been directly verified using numerical simulations~\cite{SM}.

{\em Universal anomalous statistics.---} We next address the statistical properties of the linear responses $dc_\alpha/d\gamma$, focusing on the large tail of its distribution. The latter can be predicted based on Eq.~\eqref{eq:der_LHC} and the universal DOS of soft glassy modes, $D_{\rm G}(\omega)\!\sim\!\omega^{4}$. Considering the eigen-representation of $dc_\alpha/d\gamma$ and invoking the same considerations as in~\cite{thermal_energies}, one can show that objects such as those appearing on the right-hand-side of Eq.~\eqref{eq:der_LHC} are far more sensitive to quasilocalized glassy modes than to extended phonons as $\omega\!\to\!0$ and that the $\omega$ dependence emerges only from $\calBold{M}^{-1}\!\!\sim\!\omega^{-2}$. Consequently, we have $dc_\alpha/d\gamma\!\sim\!\omega^{-8}$ and $p\left(dc_\alpha/d\gamma\right)$ is predicted to satisfy $p(dc_\alpha/d\gamma)\=D_{\rm G}(\omega)d\omega/d(dc_\alpha/d\gamma)\!\sim\!(dc_\alpha/d\gamma)^{-13/8}$ in the large $dc_\alpha/d\gamma$ limit.

\begin{figure}[ht]
 \centering
 \includegraphics[width=\columnwidth]{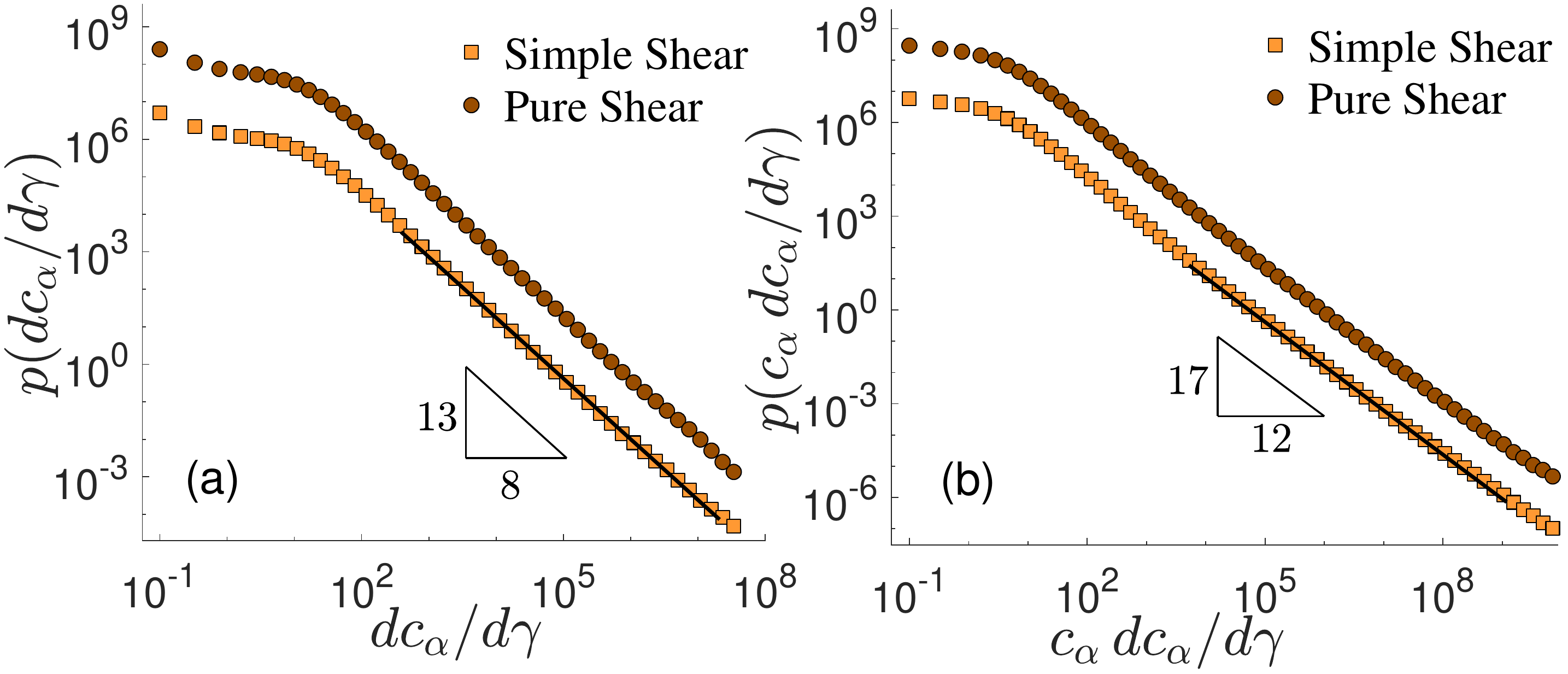}
 \caption{The probability distribution functions (a) $p(dc_\alpha/d\gamma)$ and (b) $p(c_\alpha\,dc_\alpha/d\gamma)$ for both simple (squares) and pure (circles) shear deformation. The curves in each panel are vertically shifted one with respect to the other for visual clarity, while in fact they perfectly overlap, as expected from the initial glass isotropy. The theoretical power-law predictions are marked by the solid lines and the triangles (see text for details).}
 \label{fig:statistics}
\end{figure}

To test this prediction, we performed extensive numerical simulations of a conventional computer glass-former for both simple and pure shear~\cite{SM} and extracted the statistics of $dc_\alpha/d\gamma$. The results are presented in Fig.~\ref{fig:statistics}a and are in great quantitative agreement with the theoretical prediction. We thus conclude that $dc_\alpha/d\gamma$ attains anomalously large values described by universal fat-tailed statistics related to the universal DOS of soft quasilocalized glassy modes, $D_{\rm G}(\omega)\!\sim\!\omega^{4}$. The relation between $dc_\alpha/d\gamma$ and quasilocalized modes suggests that the {\em spatial} distribution of the former features localized structures, which will be used next to construct a generalized structural predictor in glasses.
\begin{figure*}[ht!]
 \centering
 \includegraphics[width=0.98\textwidth]{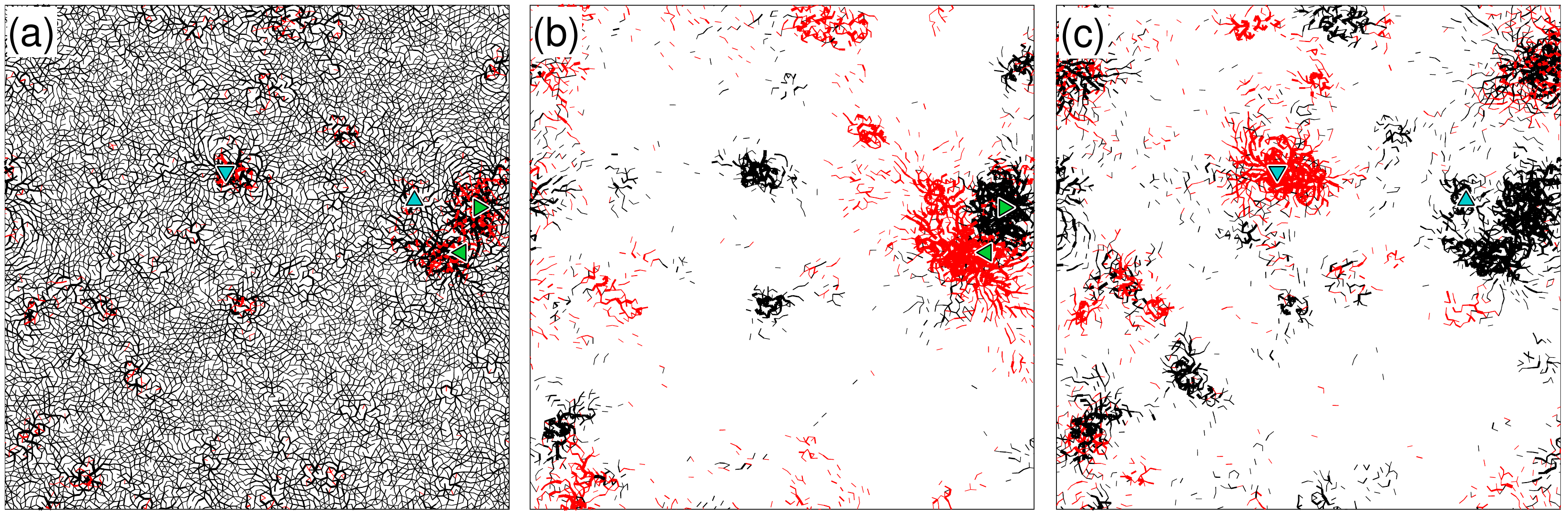}
 \caption{(a) The same as Fig.~\ref{fig:LTE}. (b) $c_\alpha\,dc_\alpha/d\gamma$ (black/red correspond to positive/negative values, line thickness represents the magnitude) for simple shear in the positive direction (the right/left triangles correspond to the first plastic events under positive/negative simple shear AQS deformation). (c) The same as panel (b), but for pure shear in the positive direction (the up/down triangles correspond to the first plastic events under positive/negative pure shear AQS deformation).}
 \label{fig:product_events}
\end{figure*}

{\em A structural predictor.---} We have at hand two quantities that appear to capture essential physical properties of soft spots in glassy materials. First, the LHC $c_\alpha$ is a scalar that quantifies the degree of softness of soft spots, i.e.~it provides a measure for how small the activation barrier for irreversible rearrangements is in some unknown direction. Second, the linear response coupling of the LHC to deformation in a certain direction $dc_\alpha/d\gamma$, which provides a measure for the degree by which externally applied forces affect the activation barrier in the direction in which they are applied. How do the two quantities combine to form a generalized anisotropic structural predictor in glasses? As both $c_\alpha$ and $dc_\alpha/d\gamma$ are predicted to attain anomalously large values at the loci of soft quasilocalized modes, we expect $dc_\alpha/d\gamma$ to single out a subpopulation of the soft spots defined by $c_\alpha$ that is most relevant for the imposed deformation in a certain direction. Consequently, we propose the product $c_\alpha\,dc_\alpha/d\gamma$ as a generalized anisotropic structural predictor in glasses.

As a first test of the idea that $c_\alpha$ and $dc_\alpha/d\gamma$ attain anomalously large values at partially overlapping locations in space, we invoke it to predict the large tail statistics of $c_\alpha\,dc_\alpha/d\gamma$. As we have $c_\alpha\!\sim\!\omega^{-4}$ and $dc_\alpha/d\gamma\!\sim\!\omega^{-8}$ in the small $\omega$ limit, the spatial overlap prediction implies $c_\alpha\,dc_\alpha/d\gamma\!\sim\!\omega^{-12}$, which leads to $p(c_\alpha\,dc_\alpha/d\gamma)\!\sim\! (c_\alpha\,dc_\alpha/d\gamma)^{-17/12}$ in the large $dc_\alpha/d\gamma$ limit (using $D_{\rm G}(\omega)\!\sim\!\omega^{4}$). This prediction is quantitatively verified in Fig.~\ref{fig:statistics}b for both simple and pure shear, lending strong support to the idea that the product $c_\alpha\,dc_\alpha/d\gamma$ indeed characterizes well-defined soft spots.

We next turn to the spatial properties of $c_\alpha\,dc_\alpha/d\gamma$, and first consider the glass realization shown in Fig.~\ref{fig:LTE}, which is shown again in Fig.~\ref{fig:product_events}a. The product $c_\alpha\,dc_\alpha/d\gamma$ under both simple/pure shear in the positive direction is shown in Fig.~\ref{fig:product_events}b-c. Here, black/red correspond to positive/negative values of $c_\alpha\,dc_\alpha/d\gamma$ (the thickness of the lines quantifies their magnitude). Two major observations can be made: (i) Soft spots that are revealed by $c_\alpha\,dc_\alpha/d\gamma$ indeed overlap those revealed by $c_\alpha$ alone, and in fact they are more pronounced (ii) There exist two subspecies of soft spots, one that is positively  coupled to deformation in a given direction (black) and one that is negatively coupled to it (red), and these subspecies depend on the direction of the deformation (cf.~panels b-c). Consequently, the product $c_\alpha\,dc_\alpha/d\gamma$ reveals orientation-dependent soft spots that offer novel predictions, which will be tested next.

{\em Quantifying the predictive power of the structural predictor.---} We first demonstrate the predictive power of $c_\alpha\,dc_\alpha/d\gamma$ using the example in Fig.~\ref{fig:product_events}; we expect plastic events to occur at one of the softest black/red spots in Fig.~\ref{fig:product_events}b when the glass undergoes simple shear deformation in the positive/negative directions, and similarly for Fig.~\ref{fig:product_events}c in relation to pure shear in the positive/negative directions. This expectation is fully supported by the results of AQS deformation simulations~\cite{SM} in the $4$ different directions, as shown by the triangles in Fig.~\ref{fig:product_events}b-c.

To systematically quantify the predictive power of the proposed structural predictor, we performed extensive computer simulations of a large ensemble of glass realizations deformed in the $4$ different directions and tracked the location of the first plastic event in each one of them. To quantify the degree of predictability, we used the following metric: the system is divided into bins of linear size $\xi\=5$ particle diameters, comparable to the localization length of soft quasilocalized modes~\cite{modes_prl_2016,modes_prl_2018}, and assigned a value obtained from the average of the structural indicator inside the bin and all of its neighbouring bins. A plastic event is assigned a rank $\lambda$ that corresponds to the fraction of the bins with a higher value than that of the bin in which it actually occurred. The best prediction corresponds to $\lambda\=0$ (the event occurred in the highest value bin) and the worst one corresponds to $\lambda\!\to\!1$ (the event occurred in the lowest value bin). When considering the cumulative distribution function $C(\lambda)$, with $0\!\le\!\lambda\!<\!1$, perfect predictability corresponds to $C(\lambda)\=\theta(\lambda)$ (Heaviside step function) and no predictability (random guess) corresponds to $C(\lambda)\=\lambda$. This metric depends on a single, physically motivated parameter $\xi$ (the quantitative dependence of the results on $\xi$ is discussed in~\cite{SM}).

\begin{figure}[ht]
 \centering
 \includegraphics[width=\columnwidth]{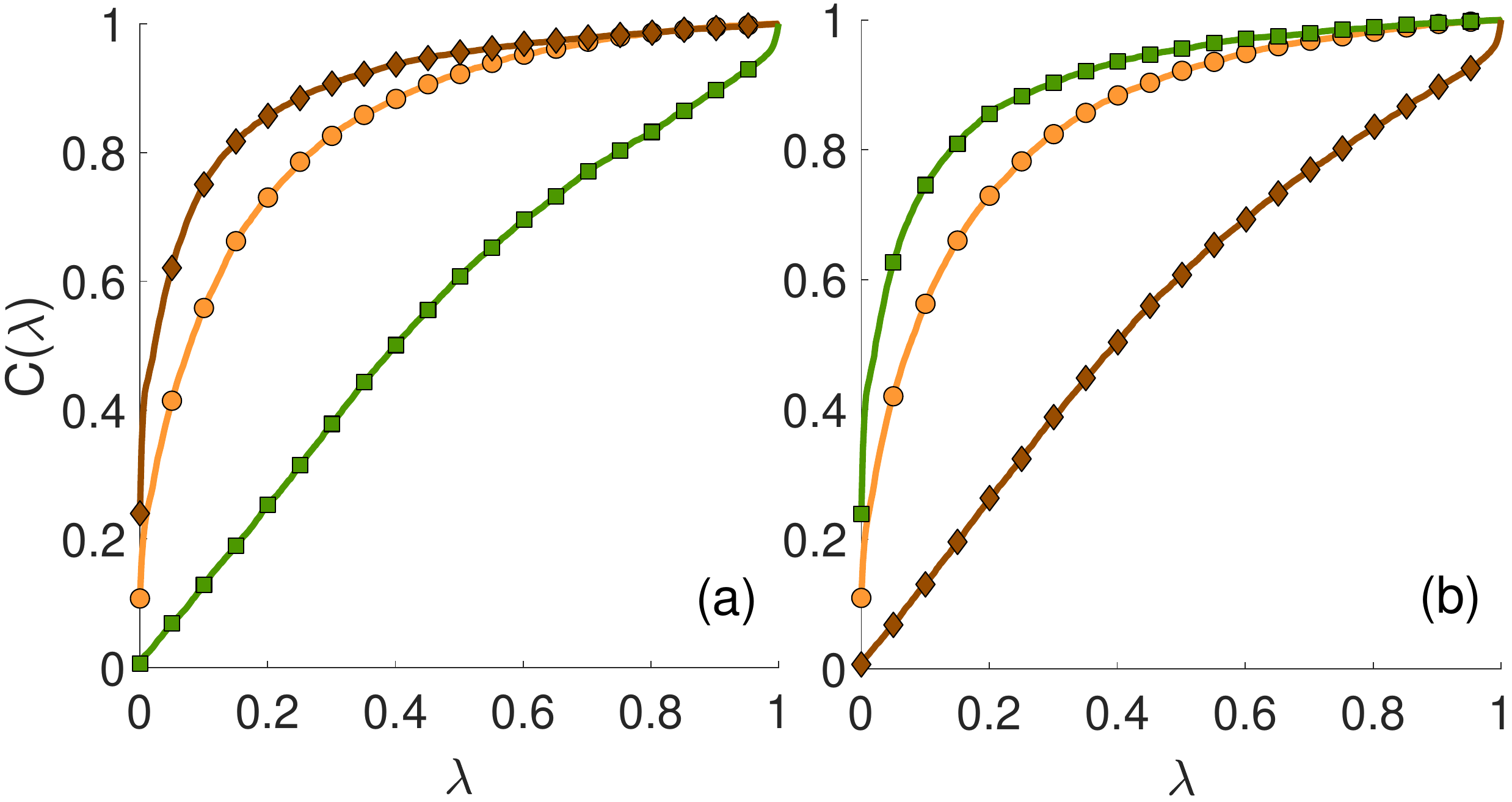}
 \caption{Quantifying the predictive power of structural indicators with respect to plastic events under AQS deformation through the function $C(\lambda)$, see text for definitions, where $C(\lambda)\!=\!\theta(\lambda)$ (Heaviside step function) corresponds to perfect predictability and $C(\lambda)\!=\!\lambda$ to no predictability. (a) Results for positive values (negative ones are set to zero) of $c_\alpha\,dc_\alpha/d\gamma$ (diamonds), for negative values (positive ones are set to zero) of it (squares) and for $|c_\alpha|$ (circles) under simple shear AQS deformation in the {\em positive} direction. (b) The same as panel (a), but under simple shear AQS deformation in the {\em negative} direction.}
 \label{fig:predictability}
\end{figure}
The results are presented in Fig.~\ref{fig:predictability}, where $C(\lambda)$ for the absolute value of the LHC $|c_\alpha|$ serves as a reference (circles). In Fig.~\ref{fig:predictability}a we consider simple shear in the positive direction, and plot $C(\lambda)$ (diamonds) for positive values of $c_\alpha\,dc_\alpha/d\gamma$ (the negative ones are set to zero). It is observed that the predictive power of $c_\alpha\,dc_\alpha/d\gamma$ is significantly larger than that of $|c_\alpha|$. $C(\lambda)$ for negative values of $c_\alpha\,dc_\alpha/d\gamma$ (the positive ones are set to zero) is also shown (squares), exhibiting essentially no predictive power, i.e.~the curve is quite close to $C(\lambda)\=\lambda$. Negative values of $c_\alpha\,dc_\alpha/d\gamma$ provide excellent predictions for plastic events once the deformation direction is reversed (that is, simple shear in the negative direction is applied), as shown in Fig.~\ref{fig:predictability}b. In fact, when the deformation direction is reversed, the black/red soft spots simply reverse their roles (while $|c_\alpha|$ remains the same, as it is independent of the direction of the driving force), as shown in Fig.~\ref{fig:predictability}b. Essentially the same results are obtained for pure shear~\cite{SM}, as expected from symmetry, further demonstrating the superior predictive power of $c_\alpha\,dc_\alpha/d\gamma$.

{\em Concluding remarks.---} The results presented above show that $c_\alpha\,dc_\alpha/d\gamma$ is a promising structural predictor in glasses. It is a first principles, model/system-independent physical quantity that reveals and highlights the orientation-dependence of soft spots inside disordered glass states. The transparent analytic structure of $c_\alpha\,dc_\alpha/d\gamma$, and its relation to quasilocalized soft excitations \cite{thermal_energies}, allows to gain physical insight into the origin of localized soft spots in glasses and their universal statistical properties. Our structural predictor involves only snapshots of non-deformed glasses and the inter-particle interactions; to the best of our knowledge, no other \emph{static} structural predictor (i.e.~that requires no dynamical information as employed e.g.~in Refs.~\cite{falk_prl_2016,Falk_prl_2018_art,Falk_local_yield_stresses_PRE_2018,machine_learning_1,machine_learning_2,machine_learning_3,machine_learning_4}) of amorphous plasticity that encodes the orientation-dependence of soft spots has been put forward before. The emerging properties of soft spots strongly echo the original Falk-Langer concept of Shear-Transformation-Zones (STZs)~\cite{falk_langer_stz} and should help developing predictive elasto-plastic models. Finally, we believe that our results offer a tool to probe the basic physics of glasses including structural relaxation, aging, memory effects and nonlinear yielding transitions.

\textit{Acknowledgments.---} E.~L.~acknowledges support from the Netherlands Organisation for Scientific Research (NWO) (Vidi grant no.~680-47-554/3259). E.~B.~acknowledges support from the Minerva Foundation with funding from the Federal German Ministry for Education and Research, the William Z.~and Eda Bess Novick Young Scientist Fund and the Harold Perlman Family. We thank J.~Zylberg for his support, advice and assistance with the numerical calculations. Z.~S.-N.~thanks Y.~Lubomirsky for useful discussions.

%

\clearpage

\onecolumngrid
\begin{center}
	\textbf{\large Supplemental Material for: ``Anisotropic Structural Predictor in Glassy Materials''}
\end{center}

\begin{center}
	{Zohar Schwartzman-Nowik$^1$, Edan Lerner$^2$, and Eran Bouchbinder$^1$}

    \textit{$^{1}$Chemical and Biological Physics Department, Weizmann Institute of Science, Rehovot 7610001, Israel\\
$^{2}$Institute for Theoretical Physics, University of Amsterdam, \\ Science Park 904, 1098 XH Amsterdam, The Netherlands}
\end{center}

\setcounter{equation}{0}
\setcounter{figure}{0}
\setcounter{section}{0}
\setcounter{table}{0}
\setcounter{page}{1}
\makeatletter
\renewcommand{\theequation}{S\arabic{equation}}
\renewcommand{\thefigure}{S\arabic{figure}}
\renewcommand{\thesection}{S-\Roman{section}}
\renewcommand*{\thepage}{S\arabic{page}}
\renewcommand{\bibnumfmt}[1]{[S#1]}
\renewcommand{\citenumfont}[1]{S#1}

\renewcommand{\dbar}{{\,\mathchar'26\mkern-12mu d}}
\newcommand{\fbar}{\bar{f}}
\newcommand{\kbar}{\bar{k}}
\newcommand{\sFrac}[2]{{\textstyle\frac{#1}{#2}}}
\newcommand{\tripleCdot}{\stackrel{\mbox{\bf\scriptsize .}}{:}}
\newcommand{\OO}{\mathcal{O}}
\newcommand\Gbo{\overline{\Gb}}

\twocolumngrid

The goal of this document is to provide additional technical details regarding the results reported on in the manuscript.

\subsection{Glass model, preparation and deformation}
\label{sec:glass_AQS}

{\bf {\em Glass model} ---} We employ a glass-forming model system in two-dimensions composed of a 50:50 binary mixture of `large' and `small' particles of equal mass $m$, interacting via radially-symmetric purely repulsive inverse power-law pairwise potentials, that follow
\begin{equation}
\varphi(r_{ij}) = \left\{ \begin{array}{ccc}\epsilon\left[ \left( \sFrac{\lambdabar_{ij}}{r_{ij}} \right)^n + \sum\limits_{\ell=0}^q c_{2\ell}\left(\sFrac{r_{ij}}{\lambdabar_{ij}}\right)^{2\ell}\right]&,&\sFrac{r_{ij}}{\lambdabar_{ij}}\le x_c\\0&,&\sFrac{r_{ij}}{\lambdabar_{ij}}> x_c\end{array} \right.,
\label{eq:IPL}
\end{equation}
where $r_{ij}$ is the distance between the $i^{\mbox{\tiny th}}$ and $j^{\mbox{\tiny th}}$ particles, $\epsilon$ is an energy scale, and $x_c$ is the dimensionless distance for which $\varphi$ vanishes continuously up to $q$ derivatives. Distances are measured in terms of the interaction lengthscale $\lambdabar$ between two `small' particles, and the rest are chosen to be $\lambdabar_{ij}\!=\!1.18\lambdabar$ for one `small' and one `large' particle, and $\lambdabar_{ij}\!=\!1.4\lambdabar$ for two `large' particles. The coefficients $c_{2\ell}$ are given by
\begin{equation}
c_{2\ell} = \frac{(-1)^{\ell+1}}{(2q-2\ell)!!(2\ell)!!}\frac{(n+2q)!!}{(n-2)!!(n+2\ell)}x_c^{-(n+2\ell)}\,.
\end{equation}
We chose the parameters $x_c\!=\!1.6, n\!=\!10$, and $q\!=\!4$. The density has been set to be $N/V\!=\!0.86\lambdabar^{-2}$; this choice sets the scale of characteristic $T\!=\!0$ interaction energies to be of order unity. This model undergoes a computer-glass-transition at a temperature of $T_g\!\approx\!0.5\epsilon/k_B$ for the chosen density.

{\bf {\em Preparation protocol} ---} We prepared an ensemble of glassy samples using the following protocol: first, systems were equilibrated in the high temperature liquid phase at $T\!=\!1.0\epsilon/k_B$. Then, the temperature was instantaneously set to a target value just below $T_g$ of the model, where the dynamics have been run for a duration $t_{\mbox{\tiny anneal}}\!=\!100\tau_0$, where  $\tau_0\!\equiv\!\lambdabar\sqrt{m/\epsilon}$ is the microscopic units of time. This short annealing step is necessary to avoid generating unphysical ultra-unstable glassy configurations that could occur in an instantaneous quench, and is computationally advantageous compared to a continuous quench at a fixed quench-rate. After the annealing step we minimized the energy to produce glassy samples by a standard conjugate gradient method. Using this protocol, we have generated $10000$ independent glassy samples, with $N\!=\!10000$. This system size has been selected as it is sufficiently large to include several soft spots, yet sufficiently small in terms of the associated computational cost.

{\bf {\em Athermal Quasi-Static (AQS) deformation} ---} The performed athermal quasi-static (AQS) deformation simulations followed well-established two-step protocols of first imposing an affine transformation $\bm{\mathcal{H}}(\gamma)$ (either simple or pure shear, in either the positive or negative directions, see manuscript for more details) to the system and then minimizing its energy while enforcing Lees-Edwards boundary conditions, see e.g.~\cite{s_lemaitre2004_avalanches,s_lemaitre2006_avalanches}. These AQS simulations have been used both to validate the analytic linear response results and to test the predictive power of the proposed structural predictor against actual plastic rearrangements/events. For the latter, the energy of the system has been used to identify plastic rearrangements/events with strain precision up to $10^{-8}$ using backtracking methods. The plastic events have been automatically spatially localized by selecting the particle with the largest displacement value as a consequence of the energy minimization step at the occurrence of a plastic event.

\subsection{Linear response coupling of the LHC to external deformation: Complete expression, numerical validation and visualization}

{\bf {\em Complete analytic expression and numerical validation} ---} As explained in the manuscript, $dc_\alpha/d\gamma$ for a given $\bm{\mathcal{H}}(\gamma)$ is obtained by operating on the LHC
\begin{equation}
c_{\alpha}\=\calBold{\varphi}_\alpha''\!:\!\calBold{M}^{-1}-{\bm f}_\alpha\!\cdot\!\calBold{M}^{-1}\cdot\calBold{U}{'''}\!:\!\calBold{M}^{-1}
\label{eq:LCH_tensorial}
\end{equation}
with the following differential operator
\begin{equation}
\frac{d}{d\gamma}=\frac{\pa}{\pa\gamma}-\calBold{U}^{\gamma}{'}\!\cdot\!\calBold{M}^{-1}\!\cdot\!\frac{\pa}{\pa{\bm x}} \ .
\end{equation}
The result takes the form
\begin{widetext}

\begin{eqnarray}
  & &\frac{dc_{\alpha}}{d\gamma} = \frac{\partial^{3}\varphi_{\alpha}}{\partial\gamma\partial x_{k}\partial x_{l}}\mathcal{M}_{kl}^{-1}-\frac{\partial^{2}\varphi_{\alpha}}{\partial x_{k}\partial x_{l}}\mathcal{M}_{kj}^{-1}\frac{\partial^{3}U}{\partial\gamma\partial x_{j}\partial x_{m}}\mathcal{M}_{ml}^{-1}-\Xi_{i}\mathcal{M}_{ij}^{-1}\frac{\partial^{3}\varphi_{\alpha}}{\partial x_{j}\partial x_{k}\partial x_{l}}\mathcal{M}_{kl}^{-1}-\frac{\partial^{2}\varphi_{\alpha}}{\partial\gamma\partial x_{k}}\mathcal{M}_{kl}^{-1}\frac{\partial^{3}U}{\partial x_{l}\partial x_{m}\partial x_{n}}\mathcal{M}_{mn}^{-1} \label{eq:lte linear response} \nonumber\\
   & &-\,\,\frac{\partial\varphi_{\alpha}}{\partial x_{k}}\mathcal{M}_{kl}^{-1}\frac{\partial^{4}U}{\partial\gamma\partial x_{l}\partial x_{m}\partial x_{n}}\mathcal{M}_{mn}^{-1} +\Xi_{i}\mathcal{M}_{ij}^{-1}\frac{\partial^{2}\varphi_{\alpha}}{\partial x_{k}\partial x_{l}}\mathcal{M}_{kn}^{-1}\frac{\partial^{3}U}{\partial x_{j}\partial x_{n}\partial x_{m}}\mathcal{M}_{ml}^{-1} \\
   & &+\,\,\frac{\partial\varphi_{\alpha}}{\partial x_{k}}\mathcal{M}_{ki}^{-1}\frac{\partial^{3}U}{\partial\gamma\partial x_{i}\partial x_{j}}\mathcal{M}_{jl}^{-1}\frac{\partial^{3}U}{\partial x_{l}\partial x_{m}\partial x_{n}}\mathcal{M}_{mn}^{-1}
    +\frac{\partial\varphi_{\alpha}}{\partial x_{k}}\mathcal{M}_{kl}^{-1}\frac{\partial^{3}U}{\partial x_{l}\partial x_{m}\partial x_{n}}\mathcal{M}_{mi}^{-1}\frac{\partial^{3}U}{\partial\gamma\partial x_{i}\partial x_{j}}\mathcal{M}_{jn}^{-1} \nonumber \\
   & &+\,\,\Xi_{i}\mathcal{M}_{ij}^{-1}\frac{\partial^{2}\varphi_{\alpha}}{\partial x_{j}\partial x_{k}}\mathcal{M}_{kl}^{-1}\frac{\partial^{3}U}{\partial x_{l}\partial x_{m}\partial x_{n}}\mathcal{M}_{mn}^{-1}
   +\Xi_{i}\mathcal{M}_{ij}^{-1}\frac{\partial\varphi_{\alpha}}{\partial x_{k}}\mathcal{M}_{kl}^{-1}\frac{\partial^{4}U}{\partial x_{j}\partial x_{l}\partial x_{m}\partial x_{n}}\mathcal{M}_{mn}^{-1} \nonumber \\
   & &-\,\,\Xi_{i}\mathcal{M}_{ij}^{-1}\frac{\partial\varphi_{\alpha}}{\partial x_{k}}\mathcal{M}_{ko}^{-1}\frac{\partial^{3}U}{\partial x_{j}\partial x_{o}\partial x_{p}}\mathcal{M}_{pl}^{-1}\frac{\partial^{3}U}{\partial x_{l}\partial x_{m}\partial x_{n}}\mathcal{M}_{mn}^{-1} -\Xi_{i}\mathcal{M}_{ij}^{-1}\frac{\partial\varphi_{\alpha}}{\partial x_{k}}\mathcal{M}_{kl}^{-1}\frac{\partial^{3}U}{\partial x_{l}\partial x_{m}\partial x_{n}}\mathcal{M}_{mo}^{-1}\frac{\partial^{3}U}{\partial x_{j}\partial x_{o}\partial x_{p}}\mathcal{M}_{pn}^{-1} \nonumber \ ,
\label{eq:full}
\end{eqnarray}
\end{widetext}
where we used the shorthand notation ${\bm \Xi}\!\equiv\!\calBold{U}^{\gamma}{'}$ for the mismatch force vector (recall that the superscript $^\gamma$ denotes the partial derivative $\pa/\pa\gamma$). Note that in the manuscript, cf.~Eq.~(2), only the leading order contribution in $\calBold{M}^{-1}$ has been reported and a compact tensorial notation has been used, while in Eq.~\eqref{eq:full} we provide the complete expression in component/index form.

The dependence of $dc_\alpha/d\gamma$ in Eq.~\eqref{eq:full} on a particular imposed deformation tensor $\bm{\mathcal{H}}(\gamma)$ is encapsulated in the  partial derivative $\pa/\pa\gamma$ that can be expressed as
\begin{equation}
\label{eq:partial_gamma}
\frac{\partial}{\partial\gamma} = \sum_{i<j}{\bm x}_{ij}\cdot\frac{d{\bm{\mathcal{H}}}^T}{d\gamma}\Big|_{\gamma=0}\cdot\frac{\partial }{\partial {\bm x}_{ij}}\ ,
\end{equation}
where ${\bm x}_{ij}$ is the inter-particle vector connecting the positions of the $i\th$ and the $j\th$ particles (using ${\bm x}_{ij}$ is natural as the pairwise potential energy $\varphi$, cf.~Eq.~\eqref{eq:IPL}, depends on the pairwise distance between interacting particles, $|{\bm x}_{ij}|\=r_{ij}$). Therefore, the dependence on the applied deformation $\bm{\mathcal{H}}(\gamma)$ is fully contained in $d{\bm{\mathcal{H}}}^T/d\gamma$ (evaluated at $\gamma\=0$).

\begin{figure}[ht!]
 \centering
 \includegraphics[width=\columnwidth]{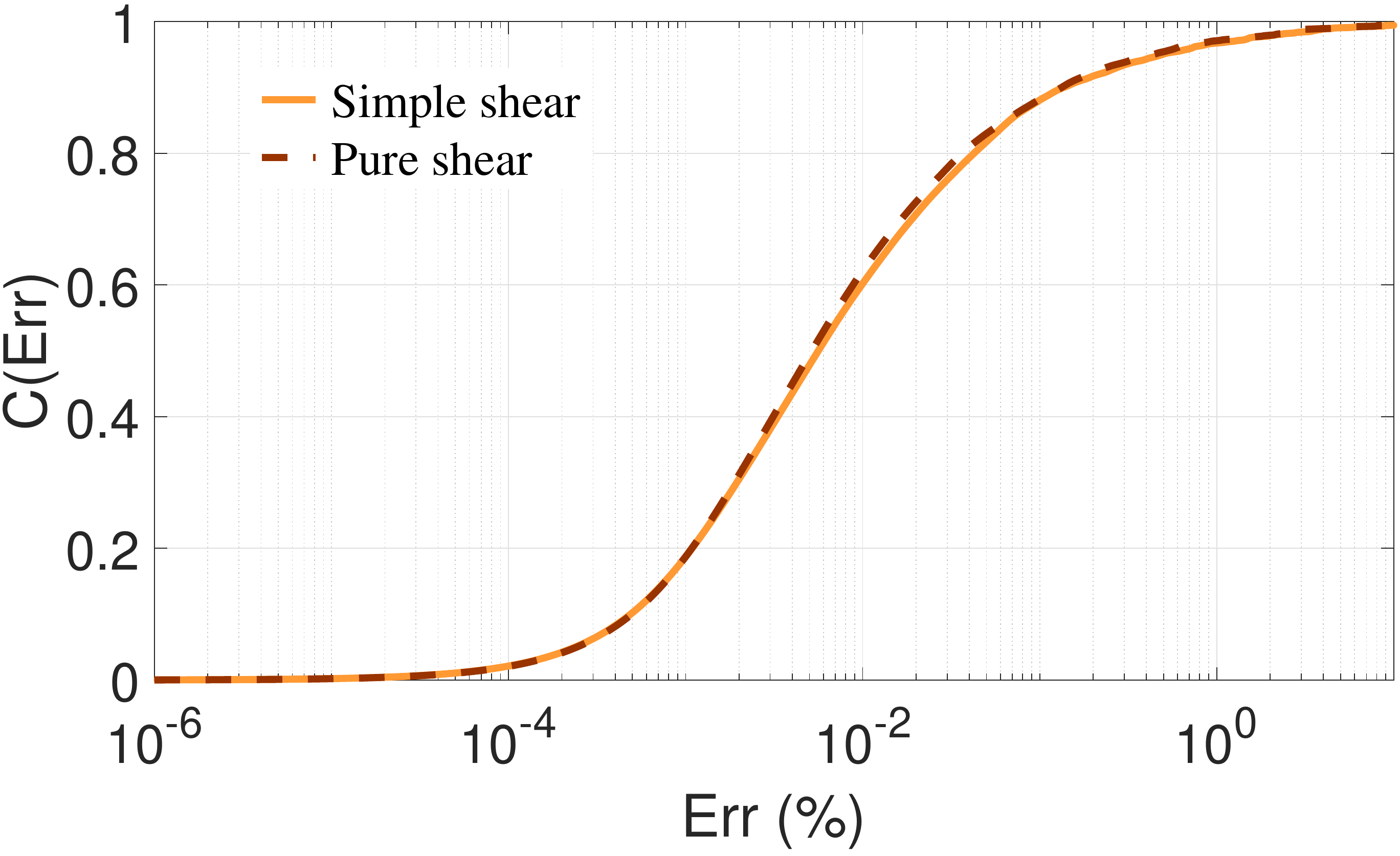}
 \caption{Numerical validation of the analytic expression for $dc_\alpha/d\gamma$ in Eq.~\eqref{eq:full} using the cumulative distribution function $C(Err)$, where $Err$ defined in Eq.~\eqref{eq:Err} quantifies the difference between $dc_\alpha/d\gamma$ and the finite differences ratio $\Delta c_\alpha/\Delta\gamma$. Data has been collected for all pairwise interactions of $1000$ glasses of size $N\!=\!10000$, for both simple (solid line) and pure (dashed line) deformation. The errors are very small, i.e.~the vast majority of pairwise interactions have an error of less than $10^{-2}$ percent, lending strong support to the validity of the analytic expression for $dc_\alpha/d\gamma$ in Eq.~\eqref{eq:full}.}
 \label{fig:numerical_validation}
\end{figure}
In order to validate the analytic expression for $dc_\alpha/d\gamma$ in Eq.~\eqref{eq:full}, we deformed $1000$ glass realizations under both simple and pure shear conditions to a strain of $\Delta\gamma\=10^{-5}$ and calculated the finite difference ratio $\Delta c_\alpha/\Delta\gamma$, where Eq.~\eqref{eq:LCH_tensorial} has been used to obtain $\Delta c_\alpha\!\equiv\!c_\alpha(\Delta\gamma)-c_\alpha(0)$. The value of $\Delta\gamma$ in the deformation simulations has been chosen to be small enough for the system to remain in the linear response regime (and in particular to avoid plastic rearrangements) and large enough for the results to be properly distinguished from the inherent noise in the calculations. In Fig.~\ref{fig:numerical_validation} we present the cumulative distribution function $C(Err)$, where the error is quantified by
\begin{equation}
Err\=100\times \frac{|dc_\alpha/d\gamma-\Delta c_\alpha/\Delta\gamma|}{\min\left(|dc_\alpha/d\gamma|,|\Delta c_\alpha/\Delta\gamma|\right)}\ ,
\label{eq:Err}
\end{equation}
for both simple and pure shear. The results, which are as expected the same for simple and pure shear, quantitatively support the validity of the analytic result for $dc_\alpha/d\gamma$ in Eq.~\eqref{eq:full}.

{\bf {\em Visualization of interaction-wise fields} ---} In Figs.~1 and 3 in the manuscript we present visualizations of interaction-wise fields of both the LHC $c_\alpha$ and of the products $c_\alpha dc_\alpha/d\gamma$. The visualization has been carried out as follows: given an interaction-wise field $x_\alpha$ ($\alpha$ is an interaction index, and $x_\alpha$ represents either $c_\alpha$ or $c_\alpha dc_\alpha/d\gamma$), we first rescale the field such that the average of the absolute magnitude of the field values is unity. We then threshold the rescaled field $\tilde{x}_\alpha$ by setting $\tilde{x}_\alpha\!=\! \tilde{x}_0$ for all $\tilde{x}_\alpha\!>\! \tilde{x}_0$, and discard of all interactions for which $\tilde{x}_\alpha\!<\!1$. The clearest visualization is obtained for $\tilde{x}_0\!=\!6.6$. The line widths presented in Figs.~1 and 3 of the manuscript are proportional to $\tilde{x}_\alpha$, and their color represent the sign of $\tilde{x}_\alpha$, with black (red) representing positive (negative) values. The exact same procedure as explained above was carried out for all presented interaction-wise fields; the differences between Fig.~3a and Figs.~3b-c stem from the different forms of the respective distributions of the different observables presented.

Finally, we note that while in the example presented in Figs.~1 and 3 the first plastic events in the $4$ different deformation directions occurred at different soft spots, this is not always the case. Yet, the systematic quantitative analysis presented in Fig.~4 in the manuscript for a large ensemble of glass realizations shows that Figs.~1 and 3 demonstrate a robust statistical effect.
\begin{figure}[ht!]
 \centering
 \includegraphics[width=\columnwidth]{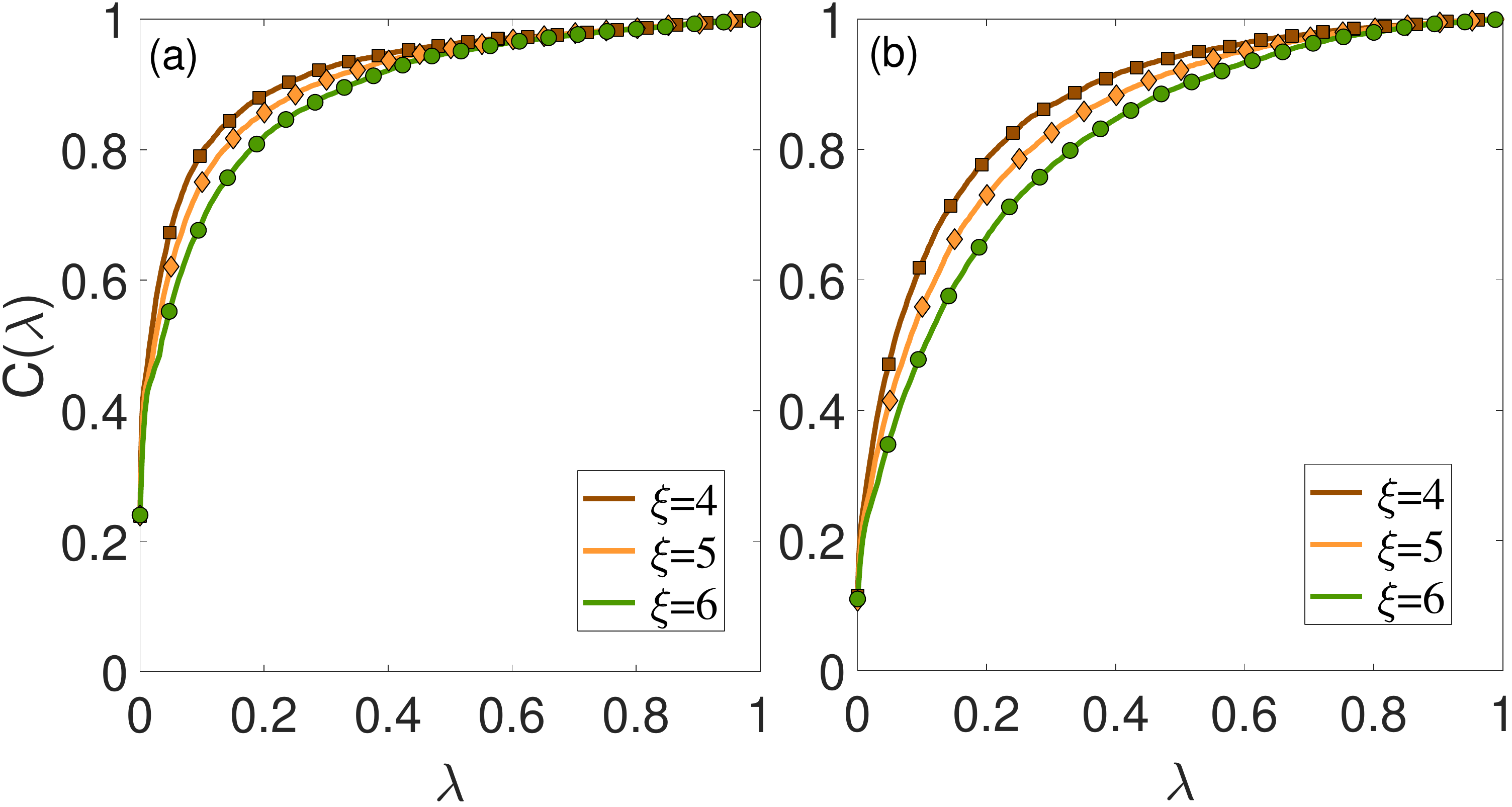}
 \caption{The bin size $\xi$ dependence of $C(\lambda)$, defined in the manuscript and plotted there in Fig.~4, for both (a) $c_\alpha dc_\alpha/d\gamma$ (under simple shear deformation in the positive direction) and (b) $c_\alpha$. The bin sizes shown are $\xi\!=\!4$ (brown line and squares), $\xi\!=\!5$ (orange line and diamonds, these results are identical to those presented in Fig.~4 in the manuscript), and $\xi\!=\!6$ (green line and circles). The predictive power of both $c_\alpha dc_\alpha/d\gamma$ and $c_\alpha$ somewhat depends on $\xi$, and appears to improve with decreasing it in the range considered here. Note that no attempt has been made in this work to optimize the predictive power with respect to $\xi$.}
 \label{fig:bin_size}
\end{figure}

\subsection{The predictive power of the structural predictor: Bin size effect and results for pure shear}

{\bf {\em Bin size effect} ---} As explained in the manuscript in detail, the metric we used to quantify the predictive power of structural indicators depends on a single, physically meaningful parameter $\xi$ that represents the typical linear size of soft spots. The latter can be estimated from the localization length of soft quasilocalized modes~\cite{s_modes_prl_2016,s_modes_prl_2018}, which is determined by the linear size of their disordered core, roughly composed of $10$ particles in linear size. Since in our metric the value assigned to each bin is averaged over neighboring bins, the latter is consistent with a bin size of $\xi\=5$, which has been used in the manuscript. In Fig.~\ref{fig:bin_size} we quantify the effect of $\xi$ on the predictive power of both $c_\alpha dc_\alpha/d\gamma$ (panel a) and $|c_\alpha|$ (panel b), by showing $C(\lambda)$ for $\xi\=4,5,6$ (from the top curve to the bottom one, respectively, where the $\xi\=5$ results are identical to those presented in Fig.~4 in the manuscript) under simple shear in the positive direction. It is observed that $C(\lambda)$ somewhat depends on the value of $\xi$, with better quantitative results obtained for smaller values of $\xi$, and that both $c_\alpha dc_\alpha/d\gamma$ and $|c_\alpha|$ exhibit similar trends. Despite that the results appear to quantitatively improve as $\xi$ is reduced, we did not aim at optimizing the predictive power relative to $\xi$ and simply used the physically sensible choice of $\xi\=5$. We note in passing that bins of $5$ particles have also been used in~\cite{s_falk_prl_2016}.
\begin{figure}[ht!]
 \centering
 \includegraphics[width=\columnwidth]{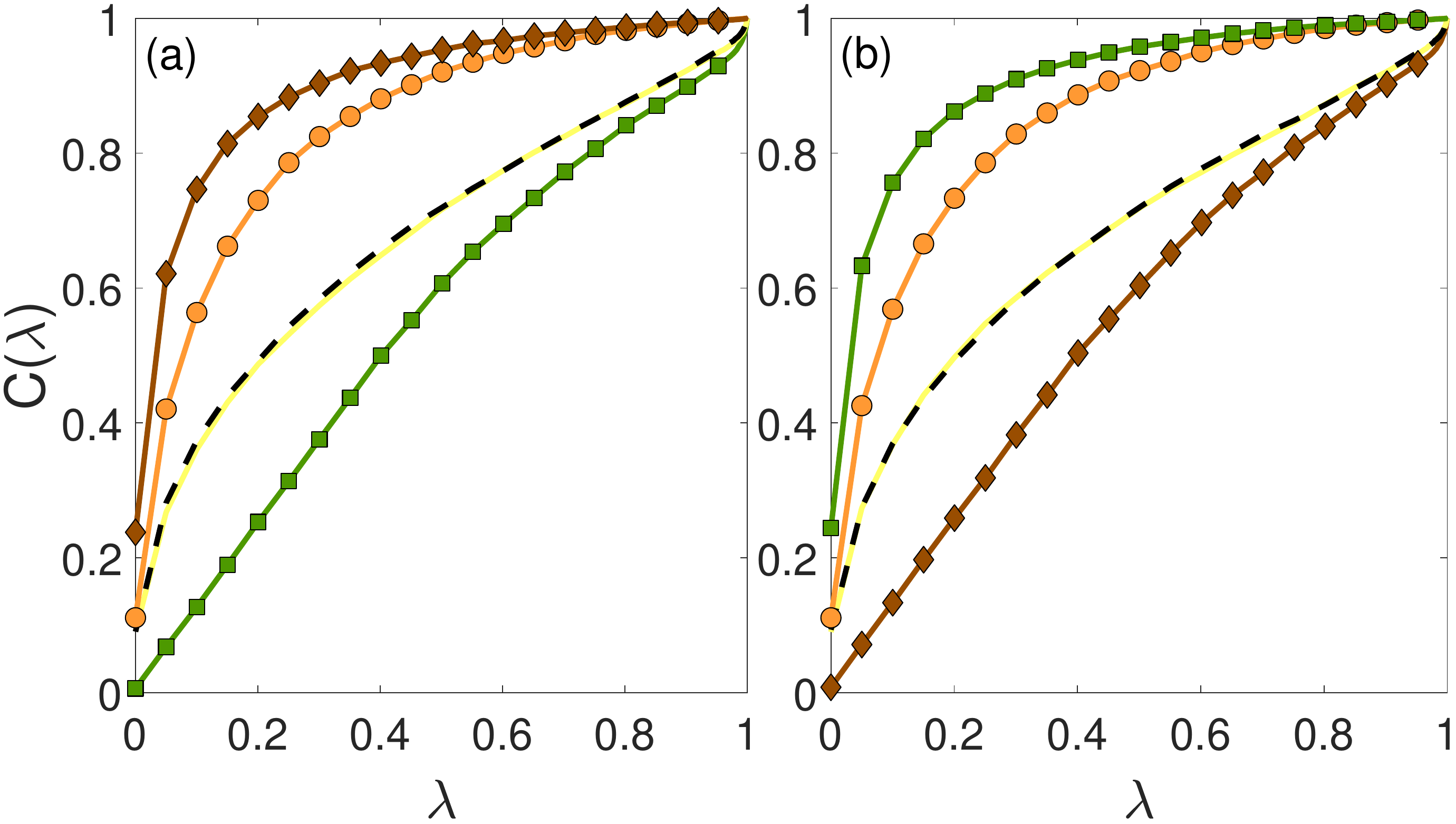}
 \caption{The same as Fig.~4 in the manuscript, where the predictability function $C(\lambda)$ is shown (see manuscript for definitions), but for pure shear in (a) the positive direction and (b) the negative one. The symbols and color code are the same as in Fig.~4 in the manuscript, but for pure instead of simple shear. In order to further highlight the orientation-dependence of $c_\alpha dc_\alpha/d\gamma$, we also added the results for $C(\lambda)$ when the glass is deformed under positive (yellow solid line) and negative (black dashed line) simple shear.}
 \label{fig:pure_shear}
\end{figure}

{\bf {\em Results for pure shear} ---} In the manuscript, results for simple shear in both the positive and negative directions have been presented in Fig.~4. For completeness, we present here the corresponding results for pure shear deformation in Fig.~\ref{fig:pure_shear}. The figure demonstrates that the results for positive and negative pure shear deformation are essentially identical for those for positive and negative pure shear deformation, as expected from the isotropy of the initial glass state.

%

\end{document}